\documentclass[12pt]{article}
\usepackage{epsfig}
\newcommand{\be}{\begin{equation}}
\newcommand{\ee}{\end{equation}}
\newcommand{\ba}{\begin{eqnarray}}
\newcommand{\ea}{\end{eqnarray}}

\renewcommand{\kappa}{{k}}

\usepackage{graphicx}

\begin{document}
\begin{titlepage}

\begin{centering}
\begin{flushright}

hep-th/0111046 \\ November 2001
\end{flushright}

\vspace{0.1in}

{\Large {\bf On the Stability of the
Classical Vacua in a Minimal SU(5)
5-D
 Supergravity Model.}}

\vspace{0.05in}

{\bf G.A. Diamandis, B.C. Georgalas, P. Kouroumalou and \\
A.B. Lahanas}
\\
{\it Physics Department, Nuclear and Particle Physics Section,\\
University of Athens,\\ Panepistimioupolis GR 157 71, Ilisia,
Athens, Greece.} \\

\vspace{0.05in}

\vspace{0.1in}
 {\bf Abstract}

\end{centering}

{\small We consider a five-dimensional
supergravity model with $SU(5)$ gauge symmetry and
the minimal field content. Studying
the arising scalar potential we find that  the
gauging  of the $U(1)_R$
symmetry of the
five-dimensional supergravity causes instabilities.
Lifting the instabilities the vacua are of
Anti-de-Sitter type and $SU(5)$
is broken along with 
supersymmetry. Keeping the $U(1)_R$ ungauged the
potential has flat directions along which supersymmetry
is unbroken.}

\vspace{0.1in}
\begin{flushleft}
\end{flushleft}

\end{titlepage}

\section{Introduction}
\indent

It is well established that the Standard Model (SM)
 describes succesfully all particle interactions at low energies.
 On the other hand it is understood that SM is an
 effective theory. At  higher energies, the 
 description of the elementary particle interactions demands
 a generalization of the SM. Assuming a unified
 description
 in terms
 of a renormalizable field theory, up to very high
 energies, lead to favorable generalization namely GUT theories
  \cite{gut1}, among which supersymmetric GUTs \cite{gut2}
  play a central r\^ole.
 Consistent inclusion of gravity dictates that these
 generalizations should be effective descriptions
 of a more fundamental underlying theory and
 String Theory \cite{string, pol}
 is the most prominent candidate
 for this aim. Indeed from the 10 dimensional
 field theory, which is the effective point limit
 of the String Theory  we can get, by suitable
 compactifications of the extra dimensions, consistent
 four-dimensional models compatible with the SM
 \cite{dienes1}.
 Along these lines it has been conjectured that
 one or two dimensions  may be compactified
 at different scales, lower from the remaining ones
  \cite{large1}. Also, after the developments
 concerning the
 duality symmetries of String Theory and in the
 framework of M-Theory \cite{wittenm, bachasm}, the
 idea that our world may be a
 brane embedded in a higher dimensional space has
 recently
 attracted much interest and has been studied intensively
 \cite{horava, lykken, tye, large2,  ovrut, kakutye, antoniadis1}.
 Besides the original compactifications new
 possibilities have been proposed \cite{randall1}. It has
 been also
 recognized
 that String/M theory may lead to brane-world models in
 which one of the extra dimensions can be even non-compact
 \cite{randall2, rubakov}.
 In all these models the four-dimensional world is
 a brane, on
 which the matter fields live, while gravity, and in some interesting
 cases
 the gauge and the Higgs fields, 
 propagate also in the transverse extra dimensions of the bulk space.

 In the majority of the cases studied, in an attempt
 to build realistic models,
 the bulk is a
 five-dimensional space \cite{peskin}.
 In these models the corresponding
 backgrounds may be of Minkowski or Anti-de-Sitter type.
 Effects of the above considerations in
 specific 
 GUT models have been
 also considered. The assumed background in these models is of
 Minkowski type and questions regarding the unification 
 and supersymmetry breaking scales
 have been addressed to. In this direction,
 assuming the fifth dimension very large, of the
 TeV scale, non supersymmetric extensions of the SM even
 without the need of unification have been considered
 \cite{dudas, dudas2, kiritsis, kawamura, altarelli, russell}.
 On the other
 hand, models embedding the SM in an Anti-de-Sitter five
 dimensional space have been also discussed
 \cite{pomarol1, pomarol}.

 In view of the aforementioned developments the study of the
 five
 dimensional supergravities has been
 revived \cite{fivesugra, gunaidin, recent, fre, guzag, ellis}.
 This is quite
 natural since after all gravity is in the center
 of all these attempts and it is legitimate to
 assume that we have to treat the fifth
 dimension before going to the "flat limit". In the 
 framework of the five-dimensional supergravity
 no specific model based on a
  particular gauge group has been studied
 so far.

 In this work we consider five dimensional supergravity
 with the minimal unified gauge group, $SU(5)$.
  The gauge symmetry enters
 with a minimal number of multiplets following the
 construction of \cite{gunaidin}. We study the emerging
 potential and discuss the Minkowski or
 Anti-de-Sitter character of the classical
 vacua. We comment on the
 fate of the gauge symmetry and supersymmetry of these vacua,
 ignoring at this stage the inclusion of matter
 fields, which do not play any r\^ole in this
 consideration.

\section{Setting the model}
\indent

We consider a five-dimensional gauged supergravity model
with $SU(5)$ gauge symmetry.  The field content of the model
is \cite{fivesugra}
 \be
 \{e_{\mu}^{m},\Psi_{\mu}^{i}, A_{\mu}^{I}, B_{\mu \nu}^{M},
 \lambda^{i \tilde{a}}, \varphi^{\tilde{x}} \}
 \label{multiplets}
 \ee
where
 \ba
 I &=& 0,1,\ldots, n \nonumber\\
 M &=& 1, \ldots, 2m \nonumber\\
 \tilde{a} &=& 1, \ldots , \tilde{n} \nonumber\\
 \tilde{x} &=& 1, \ldots , \tilde{n} \nonumber
 \ea
 with $\tilde{n}=n+2m$. The supergravity multiplet consists
 of the f\"{u}nfbein $e_{\mu}^{m}$, two gravitini
 $\Psi_{\mu}^{i}$ and the graviphoton $A_{\mu}^{0}$, where
 $i=1,2$ is the symplectic
 $SU(2)_{R}$ index. Moreover, there exist 25
 vector multiplets, $n=(a,s)$, $a=1, \ldots , 24$,
  including the
 $SU(5)$ gauge fields ($A_{\mu}^a$) plus one $SU(5)$ singlet,
labelled by $"s"$, which is
 introduced for reasons that will be explained later.
 The field content is completed by ten tensor multiplets,
 $M=1,\ldots,10$, transforming as $5+\bar{5}$ under $SU(5)$.
 The spinor and the scalar fields included in the vector
 and tensor multiplets are collectively denoted by
 $\lambda^{i \tilde{a}}$, $\varphi^{\tilde{x}}$ respectively.
 In particular the scalar fields of the model are $\varphi^{\tilde{x}}
 = \{ \varphi^{a}, \varphi^{s}, \varphi^{M} \} $.
 The indices $\tilde{a}$, $\tilde{x}$ are flat and curved
 indices respectively of the $\tilde{n}$-dimensional manifold
 $\mathcal{M}$ which is
 parametrized by the scalar fields.
 This manifold is embedded
 in an $(\tilde{n} + 1)$-dimensional space and  is determined
 by the cubic constraint
 \be
 C_{\tilde{I} \tilde{J} \tilde{K}}
 h^{\tilde{I}} h^{\tilde{J}} h^{\tilde{K}}=1
 \label{ccon}
 \ee
 where $\tilde{I},\tilde{J},\tilde{K}=0,\tilde{x}$.
 $h^{\tilde{I}}$ are functions of the scalar fields
 defining the embedding of the manifold $\mathcal{M}$.
 $C_{\tilde{I} \tilde{J} \tilde{K}}$ are
 constants symmetric in the three indices.

 For the model at hand we choose $h^{\tilde{x}}=
 \varphi ^{\tilde{x}}$, where $\varphi^{\tilde{x}}$ are
  arbitrary and  $h^0$ is to be
 determined by
 the constraint given by eq. (\ref{ccon}).
 This choice is convenient since in this case
 the constraint  becomes
 manifestly $SU(5)$ invariant.
 Taking the constants
 $C_{\tilde {I} \tilde{J} \tilde{ K}}$
 as explained in \cite{gunaidin} the constraint reads
 \be
 (h^0)^3 - 3h^0 \left[ Tr {\bf \Phi}^2
 + \frac{1}{2} \chi ^{\dagger} \chi +
 \frac{1}{2}(\varphi ^s)^2 \right] + 4 Tr {\bf \Phi}^3 -
 3\sqrt{\frac{3}{2}} \chi^{ \dagger} {\bf \Phi} \chi=1
 \label{constraint}
 \ee
 where  ${\bf \Phi} = \varphi ^{a} {\bf K}^{a}$ ,
 $a=(1,\ldots,24)$, 
 is in the adjoint representation of $SU(5)$, in
 the form of
 a $5 \times 5$ matrix.
 ${\bf{K}}^a$ denote the $SU(5)$ generators normalized
 as $Tr({\bf K}^a {\bf K}^b)=\frac{1}{2} \delta^{ab}$.
 $\chi$ are scalars in the fundamental representation
 of $SU(5)$.
 The fields $\varphi ^{M}$ introduced before are just
 the real and imaginary  parts of the fields $\chi$.
 In particular
 \be
 \varphi ^{M}=(\varphi^{A}, \varphi ^{A+5}), \qquad
 A=1, \ldots, 5
\label{realfive}
 \ee
 with
 \be
 \chi ^{A} = \varphi ^{A} + i \varphi ^{A+5},
 \qquad
 (\chi ^{\dagger})^{A} = \chi ^{\bar{A}} =
 \varphi ^{A} - i \varphi ^{A+5}.
 \label{imfive}
 \ee

 The bosonic part of the Lagrangian which is
 $SU(5) \times U(1)_R$ symmetric is given by
 \cite{gunaidin}
 \ba
 e^{-1} {\mathcal{L}} &=& -\frac{1}{2} R - \frac{1}{4}
 \mathring{a}_{\tilde{I} \tilde{J}}
 {\mathcal{H}}_{\mu \nu}^{\tilde{I}}
 {\mathcal{H}}^{\tilde{I} \, \mu \nu}
 - \frac{1}{2} g_{\tilde{x} \tilde{y}}
 ({\mathcal{D}}_{\mu} \varphi ^ {\tilde{x}})
 ({\mathcal{D}}^{\mu} \varphi ^ {\tilde{y}}) \nonumber\\
&+& \frac{e^{-1}}{6 \sqrt{6}} C_{IJK}
\epsilon ^{\mu \nu \rho \sigma \lambda}
  \Big{\{} F^{I}_{\mu \nu} F^{J}_{\rho \sigma} A_{\lambda}^{K} +
 \frac{3}{2} g F^{I}_{\mu \nu} A_{\rho}^{J}
 (f_{LF}^{K} A_{\sigma}^{L} A_{\lambda}^{F}) \nonumber\\
 &+& \frac{3}{5} g^2 (f_{GH}^{J} A_{\nu}^{G} A_{\rho}^{H})
 (f_{LF}^{K} A_{\sigma}^{L} A_{\lambda}^{F}) A_{\mu}^{I} \Big{\}}
 \nonumber\\ &+&
 \frac{e^{-1}}{4g}\epsilon ^{\mu \nu \rho \sigma \lambda}
 \Omega_{MN} B_{\mu \nu}^{M} {\mathcal{D}}_{\rho}
 B_{\sigma \lambda}^{N} \nonumber\\
 &-& g^2 P - g_{R}^2 P^{(R)}.
 \label{action}
 \ea
 The corresponding gauge fields
 are $A_{\mu}^{a}$ and $A_{\mu}=V_{0}A_{\mu}^{0} +
 V_{s}A_{\mu}^{s}$, where $V_{0}$, $V_{S}$ are constants.
 The option in which only the $SU(5)$ is gauged is
 implemented by putting $g_R=0$. In this case the
 $"s"$ multiplet is redudant.
 All scalars are neutral under $U(1)_{R}$ while under
 $SU(5)$ we have an adjoint, a $(5+ \bar{5})$
 and a singlet. The covariant derivatives appearing in the
 Lagrangian are both general coordinate  and
 gauge covariant. The fields
 ${\mathcal{H}}_{\mu \nu}^{\tilde{I}}$ describe collectivelly
 the field strengths of the vector fields and the self-dual
 tensor fields
 \be
 {\mathcal{H}}_{\mu \nu}^{\tilde{I}}=(F_{\mu \nu}^{I},
 B_{\mu \nu}^{M}).
 \ee
 The constants $f_{IJ}^{K}$ denote the structure constants
 of the gauge group. $f_{ab}^{c}$ are the
 usual $SU(5)$ structure constants and $f_{IJ}^{K}=0$
 if one of the indices is $0$ or $s$, satisfying thus
 the condition $V_{I} f_{JK}^{I}=0$. $g$ and
 $g_{R}$ are the $SU(5)$ and the $U(1)_{R}$
 gauge coupling constants respectively.
 
 The tensor $\Omega_{MN}$ in the Lagrangian (\ref{action}),
  is the given by the
 matrix
 \be
 \Omega= \left[ \begin{array}{cc}
           0 & I_{5 \times 5} \\
           -I_{5 \times 5} & 0 \end{array} \right].
\label{omega}
\ee
As far as the tensors appearing in the kinetic terms are
concerned, $\mathring{a}_{\tilde{I} \tilde{J}}$ is the
restriction of the metric of the $(\tilde{n} +1)$-
dimensional space on the $\tilde{n}$-dimensional manifold
of the scalar fields and is given by:
\be
\mathring{a}_{\tilde{I} \tilde{J}} =
 -2 C_{\tilde{I} \tilde{J} \tilde{K}}h^{\tilde{K}} +
 3 h_{\tilde{I}} h_{\tilde{J}}
\label{alpha}
\ee
where
\be
h_{\tilde{I}} = C_{\tilde{I} \tilde{J} \tilde{K}}
h^{\tilde{J}}h^{\tilde{K}}=
\mathring{a}_{\tilde{I} \tilde{J}} h^{\tilde{J}}.
\label{dual}
\ee
In eq. (\ref{action}) $g_{\tilde{x} \tilde{y}}$
is the metric of the
$\tilde{n}$-dimensional manifold ${\mathcal{M}}$
given by
\be
g_{\tilde{x} \tilde{y}} = h_{\tilde{x}}^{\tilde{I}}
h_{\tilde{y}}^{\tilde{J}} \mathring{a}_{\tilde{I} \tilde{J}}
\label{gmetric}
\ee
where  $h_{\tilde{x}}^{\tilde{I}}=-\sqrt{\frac{3}{2}}
h^{\tilde{I}},_{\tilde{x}} $. Note also that the following
relations hold
\be
h^{\tilde{I}} h_{\tilde{I}}=1, \quad
 h_{\tilde{x}}^{\tilde{I}} h_{\tilde{I}}=h^{\tilde{I}}
 h_{\tilde{I} \tilde{x}}=0
\label{relations}
 \ee
 with $h_{\tilde{I} \tilde{x}} = \sqrt{\frac{3}{2}}
 h_{\tilde{I}},_{\tilde{x}}$.

 Supersymmetry invariance requires the existence
 of additional potential terms in the Lagrangian. In particular the
 $U(1)_{R}$ gauging gives rise to the potential
 $g_{R}^2 P^{(R)}$ where
 \be
 P^{(R)}=-(P_0)^2 + P_{\tilde{a}} P^{\tilde{a}}
 \label{abel}
 \ee
 with
\be
 P_0=2h^{I}V_{I}, \qquad
 P^{\tilde{a}}= \sqrt{2}h^{\tilde{a} I}V_{I}.
\label{abeldef}
\ee
 Furthermore the existence of tensor multiplets leads to the
 appearence of $g^2P$ where
 \be
 P=2W^{\tilde{a}}W_{\tilde{a}}
\label{nonabel}
 \ee
 with  $W^{\tilde{a}}$ given by
 \be
 W^{\tilde{a}}=-\frac{\sqrt{6}}{8}h^{\tilde{a}}_{M}
 \Omega^{MN} h_{N}.
\label{nonabeldef}
 \ee
$\Omega^{MN}$ is the inverse of $\Omega_{MN}$
given in (\ref{omega}).
The conversion of flat indices ($\tilde{a}$) to curved ones
 ($\tilde{x}$), and
 vice-versa, is implemented
  by the use of $f_{\tilde{x}}^{\tilde{a}}$,
 the vielbein of the manifold
 $\mathcal{M}$.

For the specific model the quantities $h_{\tilde{I}}$
are given by
\ba
h_0  &=& (h^0)^2- \left[Tr {\mathbf{\Phi}} ^2
         +\frac{1}{2} \chi ^{\dagger} \chi +
         \frac{1}{2}(\varphi^{s})^2 \right] \nonumber\\
h_s  &=& -h^0 \varphi ^{s} \nonumber\\
h_a  &=& -h^0 \varphi ^{a} + 4 Tr {\mathbf{K}}^{a}
{\mathbf{\Phi}} ^2 - \sqrt{\frac{3}{2}} \chi^{\dagger}
{\mathbf{K}}^a \chi  \nonumber\\
h_A  &=& -\frac{1}{2} h^0 \chi^{\bar{A}} -
\sqrt{\frac{3}{2}} (\chi ^{\dagger} {\mathbf{\Phi}})_A \nonumber\\
h_{\bar{A}}  &=& -\frac{1}{2} h^0 \chi^{A} -
\sqrt{\frac{3}{2}} ({\mathbf{\Phi}} \chi)_{\bar{A}}.
\label{duals}
\ea
The metric can be written as (see eq. (\ref{alpha}))
\be
\mathring{a}_{\tilde{I} \tilde{J}} = b_{\tilde{I} \tilde{J}}
+ 3 h_{\tilde{I}} h_{\tilde{J}}
\label{alphas}
\ee
with
\be
\begin{array}{cc}
\begin{array}{ccc}
b_{00} &=& -2h^0 \\
b_{0a} &=& \varphi ^a \\
b_{0A} &=& \frac{1}{2} \chi^{\bar{A}} \\
b_{0 \bar{A}} &=& \frac{1}{2}\chi^A \\
b_{0s} &=& \varphi ^s
\end{array}
& \begin{array}{ccc}
b_{ss} &=& h^0 \\
b_{ab} &=& h^0 \delta^{ab} - 4 Tr\left[\{{\mathbf{K}}^a,
{\mathbf{K}}^b \} {\mathbf{\Phi}} \right] \\
b_{aA} &=& \sqrt{\frac{3}{2}} (\chi ^{\dagger} {\mathbf{K}}^a)_A \\
b_{a \bar{A}} &=& \sqrt{\frac{3}{2}} ({\mathbf{K}}^a \chi)_{\bar{A}} \\
b_{A \bar{A}} &=& \frac{1}{2} h^0 \delta ^{A \bar{A}} +
\sqrt{\frac{3}{2}} {\mathbf{\Phi}} _{\bar{A} A}.
\end{array}
\end{array}
\label{betas}
\ee
In eq. (\ref{betas}) only the nonvanishing
components of $ b_{\tilde{I} \tilde{J}}$ are shown.
Note that at the point $h^0=1, \quad \varphi^{\tilde{x}}=0$ we
have $\mathring{a}_{\tilde{I} \tilde{J}}=
\delta_{\tilde{I} \tilde{J}}$ as required by the consistency
of the constraint (\ref{constraint}).

\section{The potential}
\indent

The part of the potential due to the
scalar fields which are non-singlets
under the gauge group and do not belong to the gauge
multilpets is, see eq. (\ref{nonabel}),
\ba
P &=& \frac{3}{16}(h_M^{\tilde{a}} \Omega ^{MN} h_N)
      (h_P^{\tilde{a}} \Omega ^{P \Sigma} h_{\Sigma}) \nonumber\\
  &=& \frac{3}{16} f^{\tilde{a} \tilde{x}}
  f^{\tilde{a} \tilde{y}} (h_{M \tilde{x}} \Omega ^{MN} h_N)
   (h_{P \tilde{y}} \Omega ^{P \Sigma} h_{\Sigma}) \nonumber\\
   &=& \frac{3}{16} g^{\tilde{x} \tilde{y}}
   (h_{M \tilde{x}} \Omega ^{MN} h_N)
   (h_{P \tilde{y}} \Omega ^{P \Sigma} h_{\Sigma}).
\label{pform}
\ea
Using the complex notation for the fields in the $(5+ \bar{5})$
representation we find that
\be
h_{M \tilde{x}} \Omega ^{MN} h_N  =
 \sqrt{\frac{3}{2}}h_{M},_{\tilde{x}} \Omega ^{MN} h_N =
 -\sqrt{6}i(\chi_{A},_{\tilde{x}} \chi_{\bar{A}}-
   \chi_{\bar{A}},_{\tilde{x}} \chi_{A})
\label{imp}
\ee
and thus the non-abelian part of the potential receives
the simple form
\be
P= -\frac{9}{8} g^{\tilde{x} \tilde{y}}M_{\tilde{x}}
M_{\tilde{y}}
\label{nondef}
\ee
where the vector $ M_{\tilde{x}}$ 
is given by
\be
M_{\tilde{x}}=(\chi_{A},_{\tilde{x}} \chi_{\bar{A}}-
   \chi_{\bar{A}},_{\tilde{x}} \chi_{A}).
\label{mdef}
\ee
From eq. (\ref{nondef}) we see that in order to
study the potential we need the inverse of the
metric $g_{\tilde{x} \tilde{y}}$. With the adopted
form of the constraint the metric takes the form
\be
g_{\tilde{x} \tilde{y}} = -\frac{3h^0}{(h_0)^2}
 \left[ h_{\tilde{x}}h_{\tilde{y}} + \frac{1}{2}
 \frac{h_0}{h^0} F_{\tilde{x} \tilde{y}} \right]
\label{metrics}
\ee
where
\be
F_{\tilde{x} \tilde{y}} = h_{\tilde{x}}b_{0 \tilde{y}} +
  h_{\tilde{y}}b_{0 \tilde{x}}  -
   h_{0}b_{\tilde{x} \tilde{y}}.
   \label{fmetric}
\ee
The inverse of the metric is then found to be
\be
g^{\tilde{x} \tilde{y}} =
 -\frac{2}{3} h_0 \left[
 F^{\tilde{x} \tilde{y}} - \frac
 {2h^0 (F^{\tilde{x} \tilde{z}} h_{\tilde{z}})
 (F^{\tilde{y} \tilde{w}} h_{\tilde{w}})}
 {h_0 + 2 h^0 (F^{\tilde{z'} \tilde{w'}} h_{\tilde{z'}}
 h_{\tilde{w'}})} \right]
\label{inverseg}
\ee
where $F^{\tilde{x} \tilde{y}}$ is the inverse of
$F_{\tilde{x} \tilde{y}}$ in (\ref{fmetric}).
In view of the above relations the non-abelian part of
the potential can be written as
\ba
P&=&\frac{3}{4} \frac{h_0}{h_0 + 2h^0(hF^{-1}h)}
 \Big{\{} h_0 (MF^{-1}M) \nonumber \\
 &+& 2h^0
 \left[(MF^{-1}M)(hF^{-1}h)-(MF^{-1}h)^2 \right] \Big{\}}.
\label{nonabelf}
\ea

The potential stemming from the $U(1)_{R}$ gauging
consists of two terms. The first is negative definite
\be
-(P_0)^2=-4 \left[ h^0V_0+h^sV_s \right]^2
\ee
and the second is positive 
\ba
P^{\tilde{a}}P_{\tilde{a}} &=& 2 (h_{\tilde{a}}^IV_I)
(h_{\tilde{a}}^JV_J) \nonumber\\
 &=&
2g^{\tilde{x} \tilde{y}} (h_{\tilde{x}}^IV_I)
(h_{\tilde{y}}^JV_J).
\ea
With our choice of coordinates $h^{\tilde{y}}_{\tilde{x}}=
-\sqrt{\frac{3}{2}}\delta_{\tilde{x}}^{\tilde{y}}$,
and using the relation (\ref{relations}), we find that
$h_{\tilde{x}}^0=\sqrt{\frac{3}{2}}\frac{h_{\tilde{x}}}{h_0}$.
So
\be
P^{\tilde{a}}P_{\tilde{a}}=3 g^{\tilde{x} \tilde{y}}
N_{\tilde{x}}N_{\tilde{y}}
\label{abelpot}
\ee
where the corresponding vectors are given by
\be
N_{\tilde{x}}= \left(\frac{h_{\tilde{x}}}{h_0}V_0-
               \delta _{\tilde{x}}^s V_s \right).
\label{ndef}
\ee
Using the form of $g^{\tilde{x} \tilde{y}}$ found
earlier the positive term of the abelian part of the potential
is given by
\ba
P^{\tilde{a}}P_{\tilde{a}} &=&
-\frac{2}{h_0 + 2h^0(hF^{-1}h)} \Big{\{}
V_0^2(hF^{-1}h)-2V_sV_0h_0(F^{s \tilde{x}}h_{\tilde{x}})
\nonumber\\
&+& V_s^2h_0 \left[h_0F^{ss} + h^0 \left(
F^{ss}(hF^{-1}h)-(F^{s \tilde{x}}h_{\tilde{x}})^2 \right)
\right] \Big{\}}.
\label{abelposf}
\ea
The full scalar potential is given by
\be
V=g^2P-g_{R}^2 \left( P_0^2 - P^{\tilde{a}}P_{\tilde{a}}
 \right)
\label{scalarpot}
\ee
and it is rather involved to deal with
analytically. Therefore we will proceed numerically
in order to study its vacuum structure.

\section{Comments on the vacua}
\indent

The basic features of our analysis of the
scalar potential may be summarized
in the following.
\newline
{\bf (i).} If the $U(1)_{R}$ symmetry is not gauged,
then we have a positive definite potential. This potential
has flat directions along $<\chi^{A}>=0$ which preserve
supesymmetry. Thus we have a class of supersymmetric Minkowski
vacua in which the gauge group may be broken only by the scalar
fields in the adjoint representation. The degeneracy
of the vacua may be lifted by radiative corrections.
If this is the case the gauge symmetry breaking occurs in the
bulk. In the opposite case we have a supersymmetric $SU(5)$
invariant vacuum.
\cite{peskin}.
\newline
{\bf (ii).} The $U(1)_{R}$ gauging induces, as already
mentioned, a negative contribution to the potential for
supersymmetry to be preserved. This contribution
 alters the situation
drastically as we shall see in the sequel.

In what follows we will study  the symmetry breaking patterns
 $SU(3) \times SU(2) \times U(1)$,
$SU(4) \times U(1)$ and $SU(2) \times SU(2) \times U(1)$
occuring since the fields in the adjoint representation
 may develop non-vanishing v.e.v.'s along the appropriate
 directions.
For the fields $\chi ^A, \,
\chi ^{\bar{A}}$ we take only $\chi ^5=
\chi ^{\bar{5}}$ to be non-zero, to avoid charge
 violation.

As a first step we consider the model without the
multiplet $(A_{\mu}^s, \lambda^{si}, \varphi^s)$.
This multiplet, unlike the rest, is not
necessary in order to construct a locally
supersymmetric Lagrangian with $SU(5)\times U(1)_R$ gauge symmetry
in five dimensions.
In this case we find that the potential has
a local maximum at the $SU(5)$ symmetric point,
but it develops instability in the directions
$\chi ^A = \chi ^{\bar{A}} = 0$. Along these
directions the positive contribution of $P$
disappears and we are left only with $P^{(R)}$.
In $P^{(R)}$ the negative term $-(h^0)^2$, with
$h^0$  the real root of the constraint which is
continuously connected to unit, dominates the positive
term. The instability is caused by the fact that this
root keeps growing as $<\varphi^a>$  grows up. Although
this sort of
 instability is not peculiar 
for these models, see for example \cite{guzag}, it
is an unpleasent feature for a realistic model.  

\par
In an attempt to lift the instability
in a minimal way we proceed to
the inclusion of the multiplet
$\{A_{\mu}^s, \lambda^{is}, \varphi^s \}$ which
is scalar
under $SU(5)$ as we have said. Let us remark
at this point that the lowest
dimensionality of the scalar manifold arising from
string theory is 35 (see \cite{pol} Vol. II, page 309). 
This is the case in the model under
consideration after the introduction of the singlet field
$\varphi^s$. The introduction of an extra $SU(5)$
inert multiplet allows for a combination
\be
V_0A_{\mu}^0+V_sA_{\mu}^s
\ee
for the   $U(1)_{R}$   gauging, and correspondingly
modifies the negative term of $P^{(R)}$  to
\be
-(V_0h^0+V_s \varphi^s)^2
\label{negative}
\ee
 with the appropriate
change in the positive term.

In this case it turns out that
we still have the instability in the
directions $\chi ^A = \chi ^{\bar{A}} = 0$. In fact
for large values of $h^0, \, \varphi^s$ the 
negative term (\ref{negative})
dominates over the corresponding positive one.
Then the $SU(5)$-symmetric point remains either  a local minimum
or turns to a saddle point.

\par
From the study of the potential we see that with this
minimal number of multiplets it is impossible to get
rid of the instability if the range of $\varphi^s$ is
infinite. A possible way to remedy the situation is
by restricting the field  $\varphi^s$ to
take values in a finite range.
For instance
solitonic solutions of finite variation (kinks) for the
scalar fields of five-dimensional supergravity
have been  long known \cite{gukink}. Recently
explicit examples of such
solutions have been worked out for the case
of supersymmetric domain wall world embedding in  the
context of five-dimensional gauged supergravity
\cite{recent, klaus}. Furthermore when 5-d supergravity is obtained
from compactification of 11-d supergravity on a smooth
Calabi-Yau manifold such neutral scalars  emerge
naturally \cite{recent, mohaupt}. On these grounds
 it is legitimate
and well justified to
allow the field $\varphi^s$ to vary in a finite range.

 From the study of the potential we find that
 for each value of $\varphi^s$ and
  $V_s$ there exists a maximum value for
  $V_0$ for which stability of the
  potential is guaranteed. 
  This means that there exist
   non-trivial combinations of $A_{\mu}^0$ and
   $A_{\mu}^s$ for the $U(1)_R$ gauging that lead to
   a stable potential.
 For instance if we
 arbitrarily fix the values of $\varphi^s$ and
 $V_s$ and choose  $V_0=0$, the absolute minimum
 of the potential occurs when all nonsinglet
 scalars, under $SU(5)$, get vanishing v.e.v.'s.
 Therefore the vacuum is at 
 the $SU(5)$ symmetric point.
The difference now is that we have a negative
cosmological constant  given by
\[
 -g_R^2V_s (\varphi^s)^2.
\]
  Varying $V_0$  we find that
$SU(5)$ breaking
  minima develop due to the contribution of
  $h_0$ in eq. (\ref{negative}) which is determined by the
  constaint.
 The   $SU(5)$
  symmetric point turns out to be either a local maximum
  or a local minimum but in the latter
  case is not the absolute minimum. We have
  found that after some
  critical value of $V_0$ the potential becomes unstable.
We point out that at the level of 5-d supergravity
the relative values
of   $V_s$ and  $V_0$ are arbitrary.
Therefore it is a matter of the
more fundamental theory to determine their values.

  In  figures 1 and 2 we display
   the potential for particular values of
   $\varphi^s$,  $V_s$ ,  $V_0$. Since we are interested
in the directions  $<\chi^{A}>=0$, only the abelian part
is plotted. 
In the figure 1 the dashed line
corresponds to the direction
    $SU(2) \times SU(2) \times U(1)$ and the solid line to
the $SU(3) \times SU(2) \times U(1)$ direction.
  The $SU(3) \times SU(2) \times U(1)$
  minimum is lower than the
   $SU(2) \times SU(2) \times U(1)$ one. 
 In the figure 2 we plot the potential in the
    $SU(4) \times U(1)$ direction.
The corresponding minimum is slightly deeper from the
 phenomenologicaly acceptable
 $SU(3) \times SU(2) \times U(1)$ vacuum. Note however
 that the  $SU(4) \times U(1)$ direction may be
 exluded from the geometry of the manifold since
 in this direction, unlike the other cases,
  the metric has singularities. Certainly
 we do not claim that the model selects in
 this case the right vacuum naturally but
 the fact that the gauge symmetry is broken as
 a consequence of the particular  $U(1)_{R}$
 gauging is  by itself  very interesting.
\begin{figure}[ht1]
\centering
\includegraphics[scale=0.9]{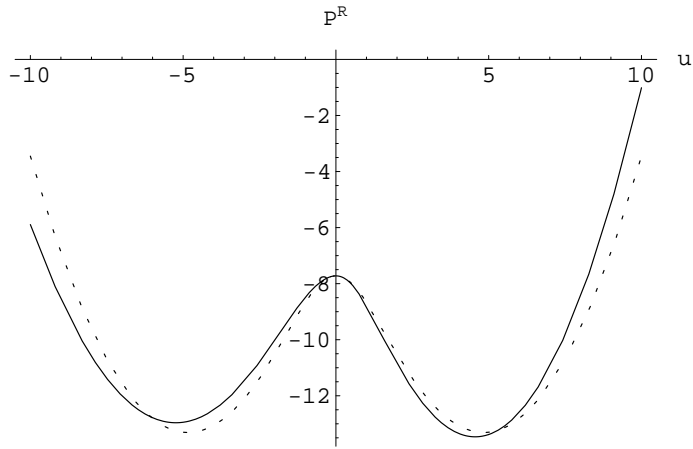}
\caption {The dashed line shows the potential in the direction $SU(2)
\times SU(2)
\times U(1)$ while the solid line shows the potential
in the direction $SU(3) \times SU(2) \times U(1) $.
 $u$ denotes the v.e.v of the $\{ \mathbf{24}\}$,
 $\varphi^a$,
in the corresponding direction.
The field $\varphi^s$ is fixed to unity. The
mixing for the $U(1)_R$ gauging is chosen as $V_s=1$,
$V_0=0.5$. At the value $V_0=0.6$ the potential becomes
unstable.}
\end{figure}
\begin{figure}[ht1]
\centering
\includegraphics[scale=0.9]{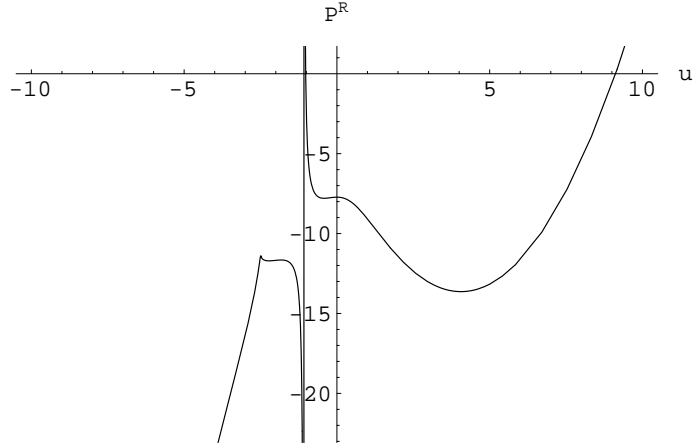}
\caption {The potential in the direction $SU(4)
\times SU(1)$. The singularity of the coordinate system
in the negative axis at $u=-1$ is shown.
The value of $V_0$, $V_s$,
$\varphi^s$ are as in figure 1.}
\end{figure}

As far as the sypersymmetric nature of the vacua
is concerned let us remind that $<\varphi^{\tilde x}>\not=0$
and $<\lambda^{i \tilde x}>= 0$ implies for the supersymmetry
transformations that  $<\delta \lambda^{i \tilde x}>= 0$.
This in turn implies that in order for
the supersymmetry
to be
preserved  
we must have $P^{\tilde a}=W^{\tilde a}=0$ when the fields 
are on their vacuum expectation values. Ignoring zero modes
of the f\"{u}nfbein the above relations imply that $M_{\tilde x}
=N_{\tilde x}=0$. However in the model at hand the
vacua which
break the gauge symmetry have $P^{\tilde a} \neq 0$
and therefore supersymmetry is broken.
Let us note  that
the vacuum expectation values of the scalar
fields enter the supersymmetry transformations of the spinors
through the combination $g W^{\tilde a} \epsilon ^i +
\frac{1}{\sqrt{2}}g_{R}P^{\tilde a} \delta^{ij} \epsilon_{j}$
 where $\epsilon_{1}= \epsilon^{2},\quad
 \epsilon_{2}=-\epsilon^{1}$. Then 
supersymmetry for the vacua
implies that
\ba
g W^{\tilde a} \epsilon ^1 &+&
\frac{1}{\sqrt{2}}g_{R}P^{\tilde a} \epsilon_{2}=0 \nonumber \\
g W^{\tilde a} \epsilon ^2 &+&
\frac{1}{\sqrt{2}}g_{R}P^{\tilde a} \epsilon_{1}=0.
\ea
So assuming that  $\epsilon^2=\epsilon^1$ the above relations
read
\be
(g W^{\tilde a}  \mp
\frac{1}{\sqrt{2}}g_{R}P^{\tilde a}) \epsilon^{1}=0.
\label{restricted}
\ee
If the vacuum expectation
values and the coupling constants are such that
either of eq. (\ref{restricted}) is  satisfied,
then we have broken supersymmetry
in the bulk space but $N=1$ supersymmetry in the
the four dimensional space.
Although very interesting this possibility is
ruled out on the grounds that a nonvanishing
value of $W^{\tilde{a}} \neq 0$ is required
since $P^{\tilde{a}} \neq 0$ too.
This adds a positive contribution to the nonabelian
part of the potential making these d=4 sypersymmetric
vacua not to be the absolute minima of the theory.
Thus in general only the $SU(5)$ symmetric point is
supersymmetric. In any other case we have simultaneous breaking
of both supersymmetry and the gauge symmetry.

 The model at hand is by itself
an effective one so it  may not be inconceivable that
$\varphi^s$ does not vary within a finite range but
the dynamics of the underlying theory drives $\varphi^s$
to a fixed value.
In this case the instability is lifted too.
A consequence of this
is that supersymmetry is completely broken. This is due to the fact
that the field $\varphi^s$ belongs to the same multiplet
with the vector field $A_{\mu}^s$ which participates
to the $U(1)_R$ gauging. Therefore the mechanism that fixes
$\varphi^s$ to a constant value, in fact sets this to be
a moduli field, is intimately connected with the
supersymmetry breaking mechanism.

In order to reach to a firm conclusion which
one of the
aforementioned
possibilities for stabilizing the potential
can be realized and exploring the supersymmetry
properties of the vacua a
 more detailed analysis 
 of the scalar field manifold is needed.
Note that the manifold ${\mathcal{M}}$ describing this model
is neither maximally symmetric nor homogeneous \cite{gunaidin}
lying  therefore outside the cases
that have been  extensively  studied
and classified so far \cite{fivesugra, gunaidin, wit}.
Besides this geometrical analysis an explicit construction
of the model from the 11-d supergravity is necessary
in order to clarify
which one of the particular vacua is favoured.

\section{Conclusions}
\indent

  We have studied the classical vacua of a five-dimensional
  $SU(5)$ supergravity model with the minimal
  field content and no matter fields. We have found that:
\newline
{\bf 1.} Without $U(1)_{R}$ gauging
the potential has flat directions determined by
vanishing v.e.v.'s of the fields in the $\mathbf{5} +
\mathbf{\bar{5}}$ representation of $SU(5)$.
Along these directions supersymmetry is preserved.
Certainly this model has to be completed with
the matter spectrum. The effective four-dimensional
model will arise either after compactification
of the fifth dimension or by considering the
corresponding action on a three-dimensional brane.
In any case the five-dimensional vacuum affects
the four-dimensional theory.
The fact that at the classical
level the gauge symmetry and the supersymmetry breaking
does not occur at the five dimensional bulk means that
such models are characterized by a large scale of
effectiveness
(energy desert \cite{bachasrev}). The possibility that
the degeneracy is lifted by five-dimensional radiative
corrections supports models
that do not exhibit unification in four dimensions.
\newline
{\bf 2.} The $U(1)_{R}$ gauging creates
stability problems. In order to
overcome the instability we introduce a vector multiplet
neutral under the gauge group. The mechanisms proposed
 are either the restriction of the variation of the
 corresponding scalar field, $\varphi^s$, in
 a finite range or fixing this field to a constant value.
 The former case may be realized
 by kink solutions existing in the context
 of five-dimensional supergravity, while
 the latter is 
 an issue related to
  the  dynamics of the
  uderlying fundamental theory. Under these
 assumptions we have found that the potential
 becomes stable in a certain range of the parameters
 $V_0$, $V_s$, which determine the mixing of the
 graviphoton and  the vector field of the
 extra multiplet which plays the r\^ole of the
 $U(1)_R$ gauge field.
 The vacua in this case are of the anti-de-Sitter type.
 As far as the fate of the gauge symmetry is concerned
 the choice $V_0=0$ yields an $SU(5)$ invariant minimum.
 In this case supersymmetry is unbroken if
 $\varphi^s$ varies in a finite range but is
 broken if $\varphi^s$ is fixed to a constant value.
 When $V_0 \neq 0$, $SU(5)$ breaking minima appear.
 In all the minima where the gauge symmetry is broken
 supersymmerty is also broken. 
  As is already quoted the
  five-dimensional vacuum determines the
  scale of effectiveness of the corresponding
  brane-world model.
For example the  $SU(5)$ breaking vacua   are  compatible
with the scenaria with low energy scale brane description.
The nonsupersymmetric  $SU(5)$ vacuum,
may be compatible
with unification with low string scale \cite{bachaslow}.

\par
Thus we see that the scenaria for physics beyond
the SM may in principle be incorporated in a unified
five-dimensional supergravity content. Certainly
the properties of the scalar field manifold,
the mechanism for stabilization of the potential
in the case of anti-de-Sitter space and the
details of including matter on the brane describing
the four-dimensional space
require further investigation. Work towards
this direction is
in progress.

\section*{Acknowledgements}
The authors acknowledge partial financial support
from the Athens University special account for research.
A.B.L. acknowledges support from
HPRN-CT-2000-00148 and
HPRN-CT-2000-00149 programmes.

\end{document}